\documentclass[10pt]{article}
\usepackage{opex2}

\newcommand{\TEM}{$\rm{TEM}_{01}\;$}

\begin{document}

\title{A high frequency optical trap for atoms using Hermite-Gaussian beams}

\author{T.P. Meyrath, F. Schreck, J.L. Hanssen, C.-S. Chuu, M.G. Raizen}
\address{Center for Nonlinear Dynamics and Department of Physics, \\ The University of Texas at Austin, Austin, Texas 78712-1081, USA }
\email{meyrath@physics.utexas.edu}

\begin{abstract}%
We present an experimental method to create a single high frequency optical trap for atoms based on an elongated Hermite-Gaussian \TEM mode beam. This trap results in confinement strength similar to that which may be obtained in an optical lattice. We discuss an optical setup to produce the trapping beam and then detail a method to load a Bose-Einstein Condensate (BEC) into a \TEM trap. Using this method, we have succeeded in producing individual highly confined lower dimensional condensates.
\end{abstract}

% The commands, \begin{OEReferences} and \end{OEReferences}
% format the References section according to OpEx standard
% style, showing the title "References".
%
% The commands, \begin{OERefLinks} and \end{OERefLinks}
% format the References section according to OpEx standard
% style, if the references also include URLs or other
% unreviewed links.  In this case the title of the section
% is "References and unreviewed links".
%
%\begin{OEReferences}

%\end{OEReferences}

\section{Introduction}
In recent years there has been much interest in development of strongly confining traps for ultra-cold atoms. Such traps are particularly attractive for the production of lower dimensional Bose-Einstein condensates (BEC). Although condensates were initially produced in magnetic traps \cite{Anderson,Davis}, optical traps are an attractive alternative due to their flexibility and may be used to generate traps of strength greatly beyond those of conventional magnetic traps \cite{footnote1}. A widely used optical trap of this sort is an optical lattice \cite{Jessen}. Optical lattice experiments have been performed in a wide variety of contexts and may be made of very high strength for the study of some quantum mechanical phenomenon \cite{Greiner,Paredes,Kinoshita,Tolra}. Lattice traps, despite this capacity, have a limitation. Typically an optical lattice has a spacing of half the laser wavelength --- order a few hundred nanometers. Condensates loaded into these traps which are initially of a size greater than $10\;\rm{\mu m}$ are typically split into thousands of distinct condensates. This limits the addressability of the individual condensates which is desirable for direct atom statistics study of some states such as the Mott-insulator transition \cite{Greiner} and the Tonks-Girardeau gas regime \cite{Paredes,Kinoshita,Tolra} and required for others such as the proposed quantum tweezer for atoms \cite{Diener}. We present an optical trap that may be used to enter these regimes, yet it a single site trap and therefore addressable.

\section{A Hermite-Gaussian optical dipole trap}
\begin{figure}
\begin{center}
\includegraphics{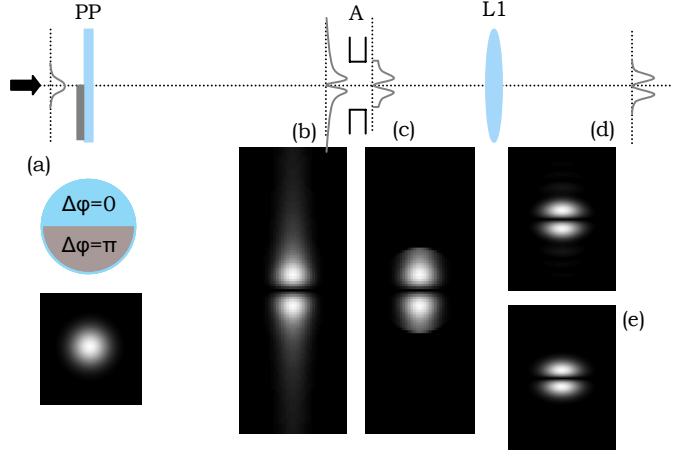}
\end{center}
\caption{Optics pictorial showing production of a \TEM from an input Gaussian. (a) An input Gaussian passes the phase plate (PP) giving a relative $\pi$ phase shift between the halves of the beam. (b) The aperture (A) is in the far field of the output beam from PP. This produces the Fourier transform at A. (c) Higher spatial modes are truncated by A. (d) Lens L1 produces the Fourier transform of the output of A resulting in a near \TEM mode profile. (e) A true \TEM beam, the profile in (d) only deviates in small fringes outside the main lobes. The images are numerically calculated beam profiles shown as $\sqrt{I(q,p)}$.} \label{fig:tem_fourier}
\end{figure}
Our approach to creating a high frequency trap for a BEC is based on a Hermite-Gaussian \TEM mode beam. This beam profile has the advantage of having a dark line in the center of the profile. For light detuned blue of an atomic transition this provides a strongly confining trap. The intensity of a Hermite-Gaussian \TEM beam is given by \cite{Saleh}:
\begin{equation}
I(q,p) = \frac{P}{\pi W_{q} W_{p}}    \frac{8q^2}{W^2_{q}} \exp\left({-\frac{2q^2}{W^2_{q}}}    {-\frac{2p^2}{W^2_{p}}}\right),  \label{eqn:TEM_equation}
 \end{equation}
in terms of the beam power $P$ and waist sizes \cite{footnote2} $W_q$ and $W_p$. The coordinates axes $q$ and $p$ are orthogonal to the axis of propagation $s$ and refer to either $x$, $y$, or $z$ depending on the beam orientation: $q=y$, $p=z$, and $s=x$ for horizontally and $q=x$, $p=z$, and $s=y$ for vertically propagating beams as in Fig. \ref{fig:beam_path}.
The optical dipole potential for a two-level atom with transition frequency $\omega_0/2\pi$ produced by light of frequency $\omega/2\pi$ with intensity profile $I(q,p)$ is \cite{Grimm,Freidman}
\begin{equation}
U(q,p) = U_0 \frac{2 {\rm e} q^2}{W^2_{q}} \exp\left({-\frac{2q^2}{W^2_{q}}}    {-\frac{2p^2}{W^2_{p}}}\right), \label{eqn:optical_potential}
\end{equation}
where the central trap depth is given by
\begin{equation}
U_0 = -{\hbar\Gamma^2\over 8 I_{\rm s}}\left({1\over\omega_0-\omega}+{1\over\omega_0+\omega}\right) {4 P\over \pi {\rm e} W_q W_p}, \label{eqn:U0}
\end{equation}
where $\Gamma$ is the natural line width of the transition, $I_{\rm s}=\pi h c\Gamma/3\lambda^3$ is the saturation intensity, and ${\rm e = exp(1)}$.
By expanding Eq. (\ref{eqn:optical_potential}) about $q=0$, the trap oscillation frequency in the $q$ direction inside the \TEM trap can be shown to be
\begin{equation}
{\omega_q\over 2\pi}=\sqrt{{\rm e} U_0\over \pi^2 m W_q^2}, \label{eqn:trap_frequency}
\end{equation}
where $m$ is the mass of the atom. From Eq. (\ref{eqn:trap_frequency}), it is clear that the trap frequency increases with larger power and smaller waist size in the $q$ direction.

A concern with using optical traps is the rate of spontaneous scattering events. Such events lead to heating of the atoms. This rate is given by \cite{Grimm}
\begin{equation}
R(q,p) = {\Gamma^3\over 8 I_{\rm s}}\left({\omega\over\omega_0}\right)^3
\left({1\over\omega_0-\omega}+{1\over\omega_0+\omega}\right)^2 I(q,p). \label{eqn:R}
\end{equation}
In a beam of intensity given by Eq. (\ref{eqn:TEM_equation}), $I(0,p)$ vanishes, but the scattering rate may be estimated by calculating it at the harmonic oscillator length $a_q=\sqrt{\hbar/m\omega_q}$. The rate is low because of the very large detuning used, typically there are no events during the optical trap time.

\section{Experimental Methods and Results}
\subsection{Hermite-Gaussian beam production}
Although it is possible to have a laser directly output a non-Gaussian beam, we use a simple holographic method to produce the \TEM mode beams which are used for the optical trap. The concept is illustrated in Fig. \ref{fig:tem_fourier}.
The input Gaussian beam passes a coated phase plate which produces a relative $\pi$ phase difference between two halves of the beam. This eliminates the Gaussian component of the beam and leaves it as some superposition of the higher modes. Because higher spatial frequencies correspond to larger angles, they are filtered by the aperture which truncates the large wings of the beam. This output is then imaged to produce a nearly \TEM profile. In Fig. \ref{fig:tem_fourier}(e) a true \TEM profile is shown along side the output of the optical system (d) which is very similar save a few small fringes.

\begin{figure}
\begin{center}
\includegraphics{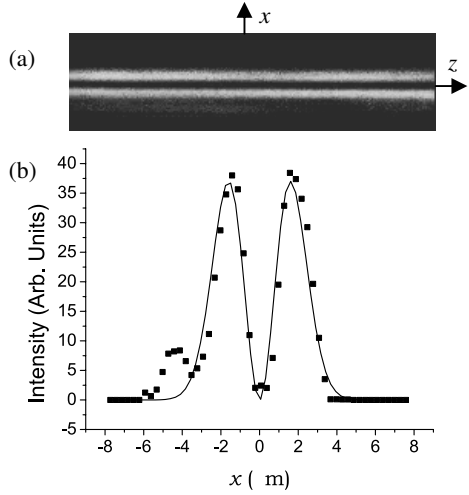}
\end{center} \caption{\label{fig:TEM01profile} Profile of \TEM mode beam. (a) A CCD picture of a \TEM trapping beam imaged as seen by the atoms. The image extends only $100\,\rm{\mu m}$ in the $z$ direction. (b) The profile along the narrow axis, the boxes are points integrated along the $z$ axis for the center $10\,\rm{\mu m}$ and the solid line is a fit to an ideal \TEM mode profile in Equation \ref{eqn:TEM_equation}, giving a waist size of $W_x=1.8\,\rm{\mu m}$. } \end{figure}
The phase plate consists of a standard BK7 double-sided anti-reflection coated window which has an additional coating on one side. The coating is a 6900$\rm{\AA}$ thick $\rm{MgF_2}$ layer that extends along half of the face with a sharp straight line boundary \cite{footnote3}.
The windows we have used were masked and coated in a vapor deposition machine.
The plate takes an input Gaussian mode beam which is aligned as to intersect the plate with half of the beam on the $\rm{MgF_2}$ coating and half off. This is the first element of the optical system shown in Fig. \ref{fig:tem_fourier}. The beam is then elongated with a cylindrical lens in the direction along the dark line to make an elliptical \TEM beam which is focused onto the atoms. A CCD image of the focus is shown in Fig. \ref{fig:TEM01profile} along with a profile in the narrow direction. The \TEM beam shown has a waist radius of 125$\,\rm{\mu m}$ in the axial direction which is much longer than the BEC region to provide a relatively uniform trap. The figure shows a fit to the ideal intensity profile of a \TEM given in Eq. (\ref{eqn:TEM_equation}) which results in $W_x=1.8\,\rm{\mu m}$.
It is noted that the measured profile fits that of an actual \TEM beam very well which justifies our use of the simple formula given by Eq. (\ref{eqn:TEM_equation}) for calculations relating to the properties of the trap itself.

\subsection{Experimental setup}
\begin{figure}
\begin{center}
\includegraphics{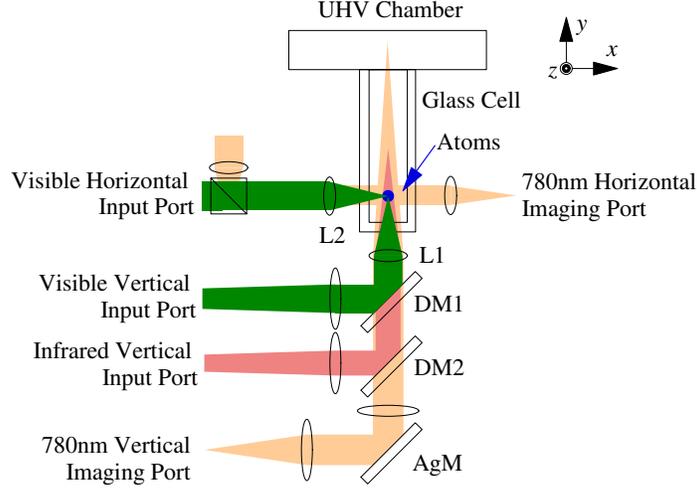}
\end{center}
\caption{Optical trap beam input ports and imaging ports in the science chamber are shown in this schematic. The setup has the capacity to accept vertical and horizontal visible beams as well as a vertical infrared beam. There are $780\;\rm{nm}$ absorption imaging beams for both vertical and horizontal diagnostics. L1 and L2 are the final lenses in the vertical and horizontal beam paths, DM1 and DM2 are dichroic mirrors, AgM is a silver mirror. Gravity in the figure is in the $-y$ direction. }
\label{fig:beam_path}
\end{figure}
For production of Bose-Einstein condensates, we have designed an experimental setup which consists of a double magneto-optic trap (MOT) system contained in an ultra-high vacuum chamber with upper and lower MOT regions separated by a differential pumping tube. The upper MOT is a six beam vapor cell MOT including an additional push beam which allows for loading of the lower MOT \cite{Wohlleben}. The lower MOT is contained by a custom ${\rm 30\,mm\times30\,mm\times115\,mm}$ (outer) rectangular fused silica glass cell with $5\,\rm{mm}$ thick walls. The glass cell is an ultra-high vacuum region (UHV) and is the location for all experiments. Surrounding the glass cell is a Quadrupole Ioffe Configuration (QUIC) Magnetic Trap \cite{Esslinger}. The experiment begins when order $10^9$ $\,^{87}$Rb atoms are loaded into the lower MOT in about $20\,$s and transferred into the quadrupole trap. The Ioffe coil is ramped on to move the atoms into a Ioffe-Prichard type magnetic trap. The trap center is offset from the lower MOT region by about $7.5\,$mm. The atoms are then cooled via forced RF evaporative cooling for about $20\,$s. This procedure produces an atom cloud of $N\approx2\times 10^6$ $\,^{87}$Rb atoms near condensation. The cloud is then transferred adiabatically into a symmetric $20\,$Hz magnetic trap in the glass cell center by ramping down the current in the quadrupole coils while keeping the Ioffe coil current high. Placing the atoms back in the center of the glass cell allows for good optical  access for the optical trap beams and for imaging of the atoms.

The beam interaction and imaging region of the setup is shown in Fig. \ref{fig:beam_path}.
The setup includes horizontal and vertical input ports for visible light, which are used for our \TEM beams. There is also an input port for a vertical infrared beam which is used for a radial confinement beam. The lenses L1 and L2 provide the final focusing of the optical trap beams. The optical trap we are describing here consists of a horizontally oriented \TEM beam (${\rm hTEM_{01}}$) which will support the atoms against gravity and a vertical infrared beam of circular gaussian profile which provides a weak radial confinement \cite{footnote4}. The trap involves several configurations which allow for loading of atoms into the h\TEM node.

\subsection{Method to load the optical trap}
\begin{figure}
\begin{center}
\includegraphics{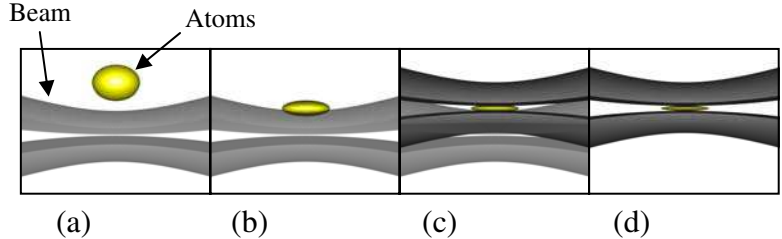} \end{center}
\caption{\label{fig:hTEM_loading}
Loading sequence of h\TEM trap. The vertical infrared beam is not shown, but present in all (b) to (d). (a) Combined optical and magnetic trap. (b) Combined gravito-optical trap where the lower h\TEM beam acts as a sheet supporting against gravity. This trap has vertical trap frequency $\omega_y \protect\cong 850\,{\rm Hz}$. (c) Transfer step into h\TEM beam. (d) Final optical trap inside h\TEM beam. This trap has vertical trap frequency $\omega_y \protect\cong 21\,{\rm kHz}$.}
\end{figure}

After initial preparation, atoms in the symmetric $20\,$Hz magnetic trap in the glass cell center may be transferred into the h\TEM beam by the steps shown pictorially in Fig. \ref{fig:hTEM_loading}.
It is not possible to directly capture the atoms into the h\TEM beam because the node spacing is only a few micrometers, but the size of the initial cloud is of order $200\;\rm{\mu m}$. The steps are as follows: (a) The h\TEM beam is ramped on below the magnetically trapped atoms and the vertical infrared beam (not shown in Fig. \ref{fig:hTEM_loading}) is turned on to provide a weak radial confinement of $60\;\rm{Hz}$. (b) The center of the magnetic trap is moved below the h\TEM beam which now acts as a sheet against which the atoms are pressed. This increases the collision rate high enough for evaporation. The magnetic trap is ramped off while the atoms evaporate radially out of the infrared trap. This produces a condensate of up to $3\times 10^5$ atoms. The atoms remain in a pancake shaped configuration pressed against the sheet by gravity. (c) The atoms are now captured inside another h\TEM which is ramped on in $200\,\rm{ms}$, $4\;\rm{\mu m}$ above the lower one.
(d) The lower h\TEM beam is then ramped off and the h\TEM beam containing the atoms is ramped up to full power.

%\begin{figure}
%\begin{center}
%\includegraphics{aom_frag.eps} \end{center}
%\caption{\label{fig:multifreq_aom}
%Acousto-optic modulator (AOM) setup driven with multiple %frequencies. For the h\TEM beam, we use an AOM with center %frequency of $110\;\rm{MHz}$ and RF bandwidth of $50\;\rm{MHz}$. It %is driven with different RF sources with independent frequency %($\nu_i$) and power ($P_i$) control added together with an RF power %combiner (PC). This results an individually controlled spot on the %imaging plane for each input frequency.}
%\end{figure}
The pair of h\TEM beams just discussed are actually both multiplexes of the same beam. The basic idea here is to use a multiple frequency acousto-optic modulator (AOM) \cite{Yariv} to produce several spots controlled independently with the driving radio frequency (RF). Such an AOM is driven with multiple frequencies of the form
$P_{\rm ac}=\sum_{i=1}^N P_i\sin^2 2\pi \nu_i t. $
In this case, $N=2$ and each signal is generated by an RF synthesizer and combined with an RF power combiner. Each frequency produces a
distinct first order diffraction angle. When imaged this leads to $N$ spots which have spacing determined by the difference in drive frequency and intensity by the RF power at each frequency. This method allows for precise control of the positions and intensities of the spots relative to each other and provides a powerful tool for condensate manipulation.

\subsection{Final trap results}
\begin{figure}
\begin{center}
\includegraphics{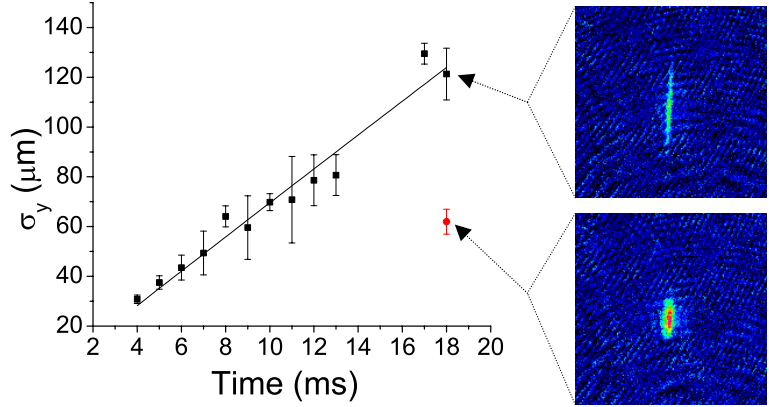}
\end{center}
\caption{\label{fig:hTEM_expansion} BEC Expansion from h\TEM trap. The black squares give data points for $\sigma_y$ as a function of time for condensates released at time zero. The data is consistent with the $21\;\rm{kHz}$ expected trap frequency for this \TEM trap. The images to the right are absorption images of one of the shots of the indicated data point. The upper picture is of a condensate released from the inside the \TEM trap, and the lower was released from above the sheet (the gravito-optical trap) as in Figure \ref{fig:hTEM_loading}(b). This is the red circle data point on the plot. The lower picture is a 3-D condenstate and does not obey Equation \ref{eqn:sigma_in_time}.  }
\end{figure}

After the final trap configuration, the atoms are released suddenly and allowed to fall. During this time, the cloud shows a change of aspect ratio which indicates that the cloud had crossed the BEC transition in the trap \cite{Castin}. The atom clouds are observed destructively with absorption imaging. This involves a $30\;\rm{\mu s}$ exposure with a resonant ($780\;\rm{nm}$) laser which is imaged onto a CCD camera.
Under the assumption that the wavefunction in the strongly confined direction is Gaussian --- that of the harmonic oscillator ground state \cite{footnote5}, the cloud image may be fit and vertical size characterized by $\sigma_y$. The rate of vertical expansion in this case depends on the trap frequency in that direction as in
\begin{equation}
\sigma_y(t)= \sqrt{2\hbar\over m\omega_y}(1+t^2\omega_y^2)^{1/2}.
\label{eqn:sigma_in_time}
\end{equation}
Condensates in a high frequency trap exhibit a much larger rate of expansion in this direction. Fig. \ref{fig:hTEM_expansion} shows expansion of atoms from a h\TEM trap. The beam used in this case had parameters: $W_y=2.4\;\rm{\mu m}$, $W_z=125\;\rm{\mu m}$, and $P=1.0\;\rm{W}$ with wavelength $\lambda=532\;\rm{nm}$, from which Equations \ref{eqn:U0} and \ref{eqn:trap_frequency} give \cite{footnote6} a trap depth of $U_0/k_{\rm B}=93\;\rm{\mu K}$ and a trap frequency of $\omega_y/2\pi=21\;\rm{kHz}$. The spontaneous scattering rate may be estimated from Eq. (\ref{eqn:R}) at the harmonic oscillator length $a_q=75\,{\rm nm}$, which results in $R(a_q,0)\cong6\times10^{-3}\,{\rm s^{-1}}$, meaning that over the time of the experiment there is unlikely to be a single event.
The line fit in Fig. \ref{fig:hTEM_expansion} indicates a trap frequency of $24\pm 4\;\rm{kHz}$, consistent with calculations.
The trap frequencies of the h\TEM trap are made high due in part to the tight focus obtained with $W_y=2.4\;\rm{\mu m}$. Because of the slightly better $f/\#$ of L1 \cite{footnote7} over L2 in Fig. \ref{fig:beam_path}, a vertical \TEM (${\rm vTEM_{01}}$) in our system can obtain even tighter focus as in Fig. \ref{fig:TEM01profile}. The largest trap frequency we have observed in our system with this method is $66\pm7\;\rm{kHz}$, which has been for this v\TEM beam with $W_x=1.8\;\rm{\mu m}$, $W_z=125\;\rm{\mu m}$, and $P=3.7\;\rm{W}$. By loading a condensate into a crossed \TEM (${\rm xTEM_{01}}$) trap consisting of both the h\TEM and v\TEM mentioned here, it is possible to create a one-dimensional BEC \cite{Meyrath}.

\section{Comments on the optical trap}
The h\TEM trap alone produces a two-dimensional condensate. The wavefunction of the condensate may be written assuming a Gaussian profile in the strongly confined direction and a Thomas-Fermi shape in the radial directions \cite{Petrov}. In the strongly confined direction, the atoms occupy the single particle ground state energy level with value $\hbar\omega_q/2$, where $2\pi\hbar$ is Planck's constant. The trap frequency, calculated from Eq. (\ref{eqn:trap_frequency}), is that for the center of the trap. Spatially, the trap frequency has a radial dependence $\omega_q=\omega_q(p,s)$ which implies a spatial ground state energy shift. This dependence is due to the Gaussian beam profile along the $p$ axis and the $W^2_{q,p}$ dependance on propagation distance $s$, see note \cite{footnote2}. The result is a weak potential of the form $m\omega_{p}^2 p^2/2$ and $m\omega_{s}^2 s^2/2$ in these directions. The radial trap frequencies may be written as
\begin{equation}
{\omega_p\over 2\pi}={i\over \sqrt{2}\pi}{1\over W_p W_q^{1/2}}\left({{\rm e} \hbar^2 U_0\over m^3}\right)^{1/4},
 \label{eqn:trap_frequency_p}
\end{equation}
and
\begin{equation}
{\omega_s\over 2\pi}=\sqrt{3\over 2}{i\over 2\pi^2}{\lambda\over W_q^{5/2}}\left({{\rm e} \hbar^2 U_0\over m^3}\right)^{1/4},
\label{eqn:trap_frequency_s}
\end{equation}
where the approximation $(W_q/W_p)^4\ll 1$ was used in deriving Eq. (\ref{eqn:trap_frequency_s}).
These complex oscillation frequencies each represent an anti-trapping potential. For the $\omega_q/2\pi= 21\,{\rm kHz}$ h\TEM trap discussed above, these anti-trapping frequencies are $\omega_p/2\pi= 12i\,{\rm Hz}$ and $\omega_s/2\pi= 40i\,{\rm Hz}$. The vertical infrared beam which provides a weak radial confinement of $60\;\rm{Hz}$ over compensates this effect and radially traps the atoms. In the case mentioned using a x\TEM configuration for a one-dimensional trap \cite{Meyrath}, $\omega_s/2\pi$ plays no role because of the geometry and only the tiny $\omega_p/2\pi$ is an issue.

Here in lies a major advantage of the \TEM trap over other blue traps, for example, a blue Gaussian sheet pair each with power $P/2$. When the sheets are separated in $q$ by $\sqrt{3}W_q$, they have an optimum trap frequency given by
\begin{equation}
{\omega'_q\over 2\pi}={1\over {\rm e}^{3/4}}\sqrt{{\rm e} U_0\over \pi^2 m W_q^2},
\end{equation}
similar to Eq. (\ref{eqn:trap_frequency}) but reduced by a factor of ${\rm e}^{-3/4}\approx 0.47$. In addition to having a lower trapping frequency in the strong direction, these traps have much larger anti-trapping frequencies. In contrast to the \TEM trap which has a zero intensity profile along the $p$ axis at $q=0$, this trap has a Gaussian axial intensity profile which produces an anti-trap in addition to the weak ground state shift anti-trap described above.
These effects result in trap frequencies
\begin{equation}
{\omega'_{p(1)}\over 2\pi}={i\over\sqrt{2}{\rm e}^{3/4}}\sqrt{{\rm e} U_0\over \pi^2 m W_p^2}\;\;\;{\rm and}\;\;\;
{\omega'_{p(2)}\over 2\pi}={i\over \sqrt{2}\pi {\rm e}^{3/8}}{1\over W_p W_q^{1/2}}\left({{\rm e} \hbar^2 U_0\over m^3}\right)^{1/4},
\end{equation}
where $\omega'_{p(1)}/2\pi$ is due to the axial intensity profile and $\omega'_{p(2)}/2\pi$ the ground state shift. These add in quadrature to give the overall $\omega'_p/2\pi$. The first does not exist for the \TEM trap and is the larger effect. The second corresponds with Eq. (\ref{eqn:trap_frequency_p}).
Using the same parameters as the h\TEM trap above, this double Gaussian trap has trap frequencies $\omega'_q/2\pi= 9.8\,{\rm kHz}$ and $\omega'_p/2\pi= 135i\,{\rm Hz}$. That is, this trap has less than half the trap frequency in the $q$ direction and more than an order of magnitude larger anti-trapping frequency in the $p$ direction.

It is worth noting that a blue trap with a dark trapping region as presented here has several advantages over a red trap \cite{Freidman}. A red trap produces an attractive potential where the atoms are high field seeking. This results in a larger spontaneous scattering rate. Because red traps use longer wavelength light, they are also more limited in focus size --- this is critical for obtaining very high trap frequencies. An additional limitation is the depth of the traps produced. If one imagines using a single red sheet used to produce a condensate as described in our h\TEM trap, it is clear that the red trap will have an attractive potential in the radial directions. Because the trap is so deep evaporative cooling is difficult.

Recent work with a \TEM beam combined with a magnetic trap has produced quasi-2D condensates \cite{Smith}. In their experiment, the authors operate in a regime where the focus in both $q$ and $p$ directions is much larger than in the case presented here. Because of the longer depth of focus and larger $W_p$ used in their experiment, the optical trap has a more uniform potential. This is a preferred regime in which to operate for the study of vortex dynamics in two-dimensions, for example, as the authors point out. The central trap frequencies, however, obtained are considerably smaller: by a factor of 10 for the h\TEM trap and a factor of 30 for v\TEM trap than in the case presented here. For the possibility of entering regimes of quantum gases as mentioned earlier \cite{Greiner,Paredes,Kinoshita,Tolra,Diener}, such trapping strengths are required.

\section{Conclusions}
We have proven a method to create \TEM beams with a holographic plate and provided a method for loading a BEC into the tightly confined \TEM trap. This method pushes the possibilities of making single lower dimensional condensates. The trapping strengths which have been obtained in these single site optical traps are comparable to those often obtained in optical lattices. This technique paves the way towards experiments with single highly confined condensates.

\section*{Acknowledgements}
The authors would like to acknowledge support from the NSF, the R.A. Welch Foundation, discussions with A.M. Dudarev and G. Price. T.P.M. acknowledges support through the NSF Graduate Research Fellowship and F.S. through the Alexander von Humboldt Foundation.


\begin{thebibliography}{}
\bibitem{footnote1} Conventional magnetic traps refer to Ioffe-Prichard type or TOP traps, atom chip setups may also generate high strength traps because of the close proximity of the atoms to the chip surface \cite{Folman}.
\bibitem{footnote3} $\rm{MgF_2}$ was chosen because it is a soft coating and more easily makes a sharp feature. This thickness is for a $\pi$ relative phase difference at wavelength $\lambda=532\,{\rm nm}$
\bibitem{footnote4} The \TEM traps discussed here use laser wavelength $\lambda=532\;\rm{nm}$ and the infrared vertical trap uses $\lambda=1064\;\rm{nm}$. These are {\sl blue} and {\sl red} relative to the strong transitions of rubidium near $780\;\rm{nm}$. Blue traps result in a repulsive potential whereas red traps result in an attractive potential \cite{Grimm}.
\bibitem{footnote5} This is what it means to be lower dimensional, the motion in the strongly confined direction is frozen out and the wave function may be written as the ground state of a single particle harmonic oscillator in the direction in question. This occurs when the three-dimensional chemical potential $\mu_{\rm 3D}$ is below the harmonic oscillator energy state splitting of the strong direction $\hbar\omega_q$ \cite{Petrov}.
\bibitem{footnote6} We use values for the ${\rm D_2}$ line of rubidium 87: $I_{\rm s}\cong 1.67\,{\rm mW/cm^2}$, $\Gamma\cong 2\pi\cdot 6.065\,{\rm MHz}$, $\omega_0\cong 2\pi\cdot 384.23\,{\rm THz}$, and $m\cong 1.443\times 10^{-25}\,{\rm kg}
$. In Eq. (\ref{eqn:U0}), it is important not to use the common rotating wave approximation (which is to assume $|\omega_0-\omega|\ll\omega_0+\omega$ and neglect the second term in parenthesis) because the detuning is too far for this to be valid. This is a larger effect than explicitly including the ${\rm D_1}$ line which is typically done for nearer detunings.
\bibitem{footnote7} This is a gradium index lens with $f/\#=2.2$.
\end{thebibliography}
\end{document}